# To quote this work :

**Please use :**
 J.S. Wu, V. Brien, P. Brunet, C. Dong and J.M. Dubois
*Scratch-induced surface microstructures on the deformed surface of Al-Cu-Fe icosahedral quasicrystals*
Materials Science and Engineering, 294-296 , 846-849 (2000)
10.1016/S0921-5093(00)01043-1, hal-02882488
**Thank you**

# Scratch-induced surface microstructures on the deformed surface of Al-Cu-Fe icosahedral quasicrystals

**J.S. WU[1,2], V. BRIEN[1,*], P. BRUNET[1], C. DONG[3] and J.M. DUBOIS[1]**

[1] Laboratoire de Science et Génie des Matériaux Métalliques (CNRS UMR 7584) and GDR CINQ, Centre d'Ingénierie des Matériaux, Ecole des Mines - Parc de Saurupt, 54042 Nancy Cedex (France)

[2] Beijing Laboratory of Electron Microscopy, Institute of Physics, Chinese Academy of Sciences, P.O. Box 2724, 100080 Beijing (P.R. China)

[*] Corresponding author : brien@mines.u-nancy.fr

[3] Department of Materials Engineering, Dalian University of Technology, 116023 Dalian (P.R. China)

**Abstract**
Scanning electron microscopy (SEM) and transmission electron microscopy (TEM) investigations of sintered Al-Cu-Fe icosahedral quasicrystal (IQC) have been carried out to understand the origin of some ductility previously noticed within tracks produced by standard tribological scratch tests. Transformation of the icosahedral phase to a modulated structure is shown and a transformation of the IQC to a bcc phase has been found beneath the tracks. Twins and dislocations have also been observed.

## 1.    Introduction

Despite their well-known brittleness at room temperature quasicrystals that limits their technological application as bulk materials, quasicrystals present very good wearing properties that make them extremely interesting to be used as coatings on soft metals. Within this frame, the authors of [1] and [2] carried out a standard scratch test in order to characterize the tribological behaviour of a polygrained Al-Cu-Fe material. On one hand, precious information was obtained on the friction and wear properties, on the other hand, the papers report that little but definite ductility was spotable within the area sheared by the loaded indenter whereas no ductility could be detected outside. In order to understand the origin of this ductility we decided to repeat the experiment and carry out a detailed microstructural investigation of the material under and outside the scratch track.

## 2.    Experimental procedures

Preparation of the sintered poly-grained $Al_{62}Cu_{25.5}Fe_{12.5}$ whose grain size is = 25-45 μm is detailed in [1] and [2]. One just precises that to make sure no residual CsCl-type cubic phase is left in the sample, an annealing treatment of 3 hours at 1083 K below the peritectic formation temperature of the icosahedral phase has been performed. This was confirmed by X-ray diffraction and transmission electron microscopy (TEM). A standard unidirectional multipass scratch test was then performed on a mirror polished face of the sintered disk with a CSEM Revetest device. The constant normal load applied on the WC-Co spherical indenter (diameter 1.58 mm) was 30 N. Five passes were performed. At this stage, optical and scanning electron microscopy (SEM) have been used to look at the surface damage produced. Thin foils have been prepared cutting the sample in order to get a plan-view of the scratches. Tripode and subsequent argon ion-milling performed on the side of the scratch allowed us to prepare a TEM specimen well positionned below the scratch. TEM observations were done on a Philips CM12 at 120kV.

## 3. Results and discussion

As it is visible on the SEM micrograph in Fig. 1, the damage caused by such a test on the quasicrystal is typical of a dry sliding test. A strong trace of the ball pass composed of multiple and parallel thin scratches has been left. Regularly spaced partial ring cracks perpendicular to the main track as well as many variably sized particles (range : 10-30nm) can be found. These particles have been left from a transfer layer known to appear between the indenter ball and the material surface during such a test and consisting of redeposited material. Indication that some plastic deformation must have occurred is proved by localised roughness of the surface. On all images of this paper a big arrow indicates the direction of the scratch.

A deeper study performed by Transmission Electron Microscopy, shows that different transformations of the initial microstructure have occurred under the scratch track. The one concerning the biggest volume of matter is an extensive modulation of the icosahedral structure (Fig. 2). Indeed Electron Diffraction Patterns (EDP) taken along a five-fold or two fold axis of the IQC structure in this modulated region (Fig. 2b and 2c) display significant differences when compared to the five or two fold patterns taken far away from the scratch (Fig. 2a). Fig. 2a is the EDP of the non-deformed IQC located outside the scratch along the five-fold zone axis. Under the scratch, the five-fold zone axis pattern (Fig. 2b) presents 1/ streaked satellites spots mainly in two sets along the two fold A2 and a projection of a three-fold axis A3' and 2/ extra satellites spots shaped either in a triangle shape (mark 1) either in a set of three aligned spots (mark 2). On the two-fold zone axis pattern still taken under the scratch, reflections along the three-fold axis present also sharp satellites spots with a local base-centered symmetry. All these extra spots indicate a kind of long-distance ordering has occurred within the structure. A dark field image (Fig. 2c) was made using a set of triangle-shaped satellites spots from the five-fold zone axis. It reveals a set of short striations whose directions are parallel to the A2 and A3' axis showing another aspect of the modulation of the icosahedral quasicrystal (IQC) structure. Such a modulation has already been encountered in previous literature. In [3] it was reported that it occurs during a reversible transformation from a rhombohedral approximant crystal to a quasicrystal $Al_{63.5}Cu_{12.5}Fe_{24}$. In the same way, Liu et al. [4] could observe a modulation in the course of a continuous decomposition of an Al-Cu-Fe IQC. The modulation state does then seem to be associated with an out of equilibrium situation which matches well the case here considering the very short time of localized loading during the scratch test. Incomplete atomic diffusion and complex phason state should be the reasons for the appearance of this modulation.

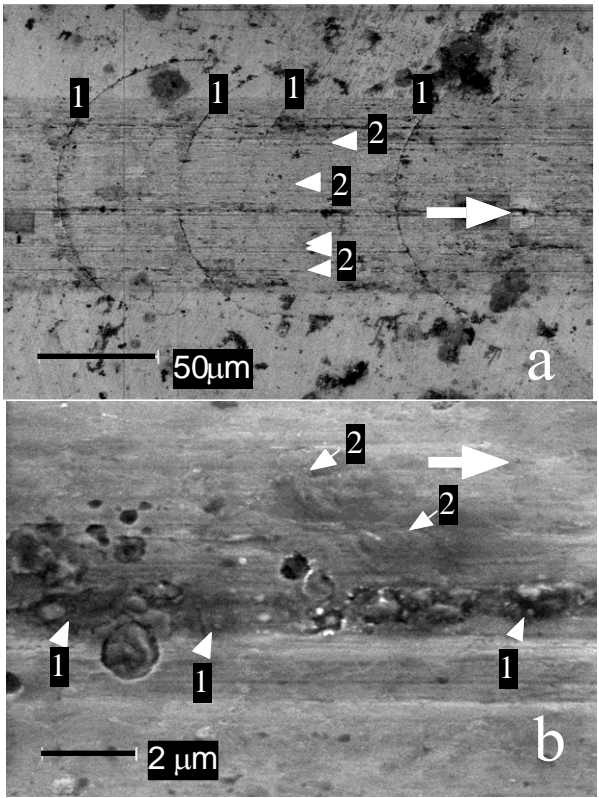

Figure 1 : Scanning Electron Microscopy picture of the scratch track. Big white arrow indicates the direction of the indenter movement. a/ Global view. Marks 1 show the partial rings generated when testing. Marks 2 spot the numerous crack lines. b/ detail of view a/. Marks 1 spot the particles of redeposited matter. Marks 2 show local roughness indicating a plastic deformation under the surface.

The second noticeable transformation is a phase transformation of the IQC structure to a bcc structure. Indeed small nanometer sized particles fully embedded within the quasicrystalline phase and aligned along the scratch track direction can be found (Fig. 3). Figure 3b,3c displays the microdiffraction pattern obtained to determine their structure. One can notice on the top of the spots stemming from the IQC structure, periodically spaced spots revealing the structure factor of a bcc crystallographic structure (lattice parameter a=0.29 nm). The later ones have been indexed and marked by asterisks. Although not identical, this phase is similar to the β-cubic CsCl-type phase which can be found in the middle part of the Al-Cu-Fe phase diagram [5]. They both display the same parameter value, but the structure factor of the CsCl-type phase provokes no spot extinction on the cubic diffraction pattern while the bcc one presents extinction for odd sums of h, k and l. Such spots are indeed present in our diffraction patterns. The determination of the orientation relationships between these particles and the surrounding IQC gives a standard result for an orientation relationship between a cubic phase and an IQC structure. A5 // <110> ; <113> and A2 // <110> ; <111> ; <112>. Modelling of the spatial stress distribution under the scratch gives the maximum stress concentration (called Hertz pressure) along a line parallel to the ball way at some distance in depth of the surface [6]. This line seems to correspond very well with the alignment of the bcc nodules. This makes us think that the bcc nodules must have been created under high stresses during the test. This IQC-bcc transformation can be compared to a transition from the Al-Cu-Fe IQC to another cubic phase ($a$=0.83 nm) mentioned by Kang et al. [7] to appear under compression. Similarly, Yu et al. [8] have found such a transformation in an icosahedral Al-Li-Cu under uniaxial compression.

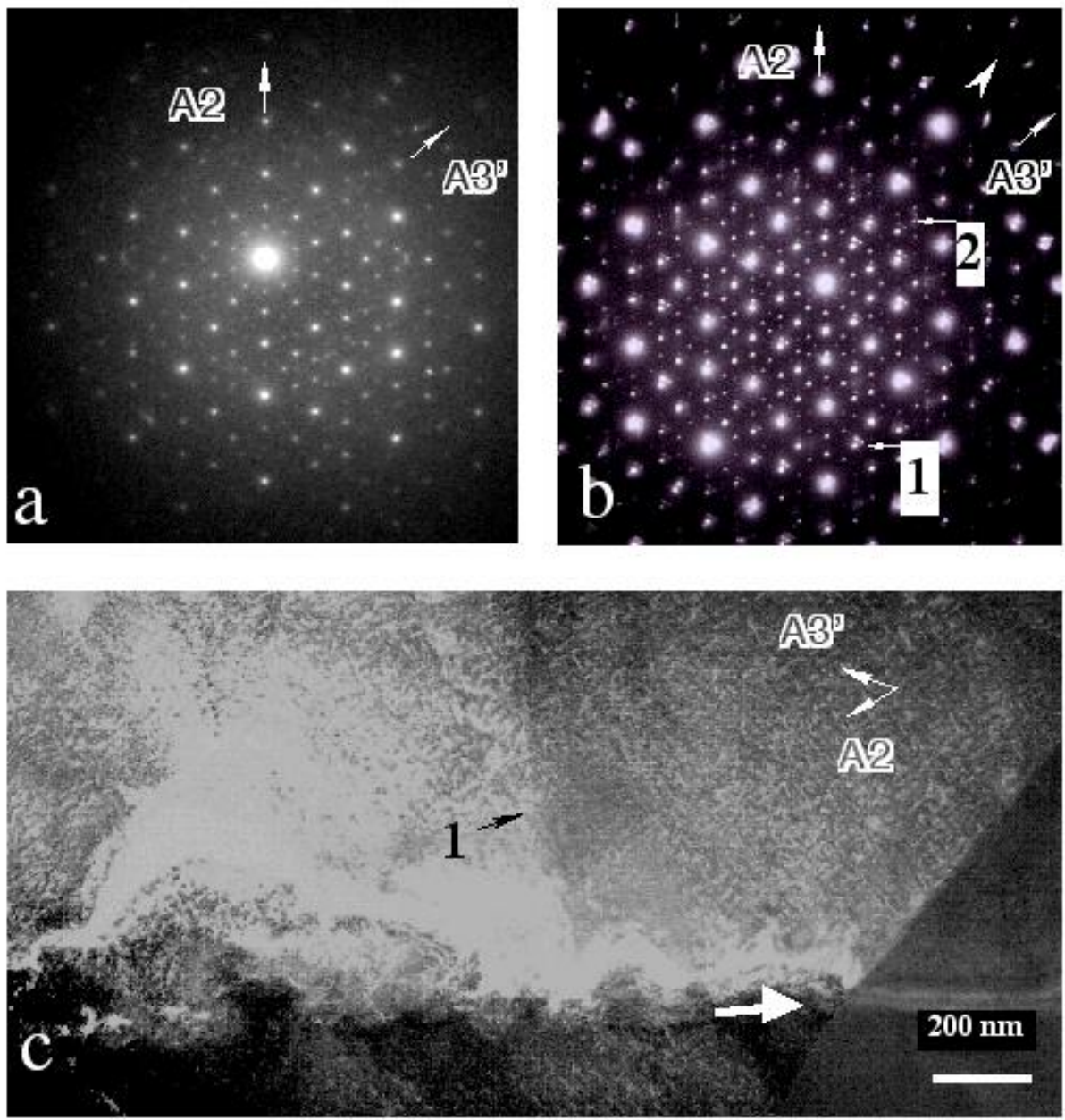

Figure 2 : Transmission Electron Diffraction Patterns and image showing the modulation of the IQC structure under the scratch. a/ Five-fold axis of the IQC far from the scratch track b/ Five-fold axis of the IQC located under the scratch track c/ Dark field image built with a triangle set of spots of EDP b/ .

In a much less significant way, other events of plastic deformation could be identified. Splitting of the grains, residual strain and even some dislocations and twins could be observed emerging form the edges of smaller scratch tracks.

As stated in the introduction, the deformation mechanisms of Al-Cu-Fe are rather unclear. Considering that the stress state under the scratch is very complex since it is a mixing of compression and shearing, the test performed here gives a good illustration of the diversity of plastic events that can happen in Al-Cu-Fe. It is quite obvious that further mechanical characterization is necessary to fully address the mechanical behaviour of this quasicrystal. However, the present microstructure observations show that the slight ductility rise macroscopically observed by the authors of [1-2] is linked to microscopic plastic deformation, and could be the result of the observed modulation of the icosahedral structure.

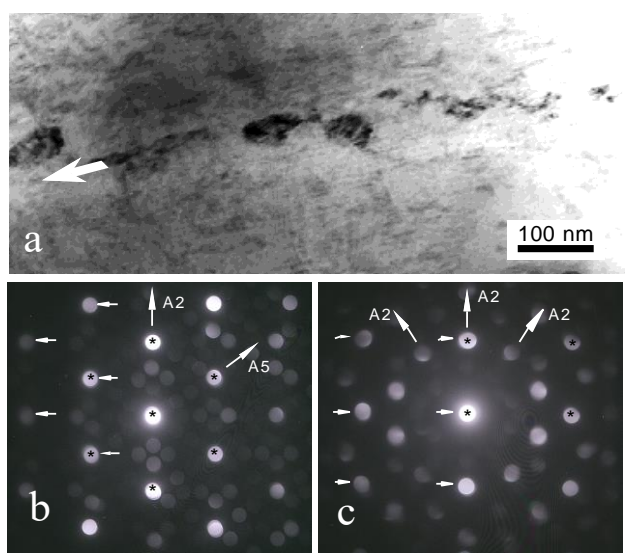

Figure 3 : a/ Transmission Electron Micrograph showing the b.c.c. nodules. b,c/ Micro-diffraction patterns of the nodules, b/ Two fold axis of the surrounding IQC, c/ Five-fold axis of the surounding IQC

## 4. Conclusion

Detailed microstructural observations have shown that the fragile Al-Cu-Fe quasicrystal can accommodate some plastic deformation during dry sliding tests performed at room temperature. SEM and TEM investigations have shown that this plasticity is linked to an extensive modulation of the IQC, a transformation to a bcc phase under the shape of nodules showing a standard crystallographic orientation relationship with the surrounding IQC phase and to some dislocations and twins that were identified under the scratch track.

## Acknowledgements
We acknowledge the financial support of the Chinese-French Advanced Research Program (PRA MX 96-02), the CNRS, the CUG-Nancy (CPER 1994-99) and the National Natural Science Foundation of China (grant No. 59525103). We thank J. von Stebut for the provision of scratch-test facility and him, C. Comte and J.P Bellot for interesting discussions. We are most grateful to G. Beck and K.H. Kuo for their constant interest in this work.

## References
[1] S.S. Kang, J.M. Dubois and J. von Stebut, *J. Mater. Res.*, 1993, 8, p. 2471
[2] J. Von Stebut, C. Strobel and J.M. Dubois, in *Quasicrystals,* eds. C. Janot and R. Mosseri, (World Scientific, Singapore), 1995, p. 704
[3] M. Audier, Y. Bréchet, M. de Boissieu, C. Janot and J.M. Dubois, *Phil. Mag.* B, 1991, 63, p. 1375
[4] W. Liu, U. Köster and A. Zaluska, *Phys. Stat. Sol.*, 1991, a 126, k9.
[5] J.M. Faudot, *Ann. Chim. Fr.*, 1993, 18, p. 445
[6] G.M. Hamilton and L.E. Goodman, Journal of Applied Mechanics, 1966, p. 371
[7] S.S. Kang, S.S. and J.M. Dubois, *Phil. Mag.* A, 1992a, 66, 151
[8] D.P. Yu, N. Baluc, W. Staiger and M. Kléman,1995, *Phil. Mag. Lett.*, 72, p. 61